# Maternal Characteristics and Newborn Birth Weight: A Comprehensive Statistical Analysis

**Authors** - Prithwiraj Chatterjee, Abhinav Tanwar, Devadharshini Udayakumar

## Abstract

This report presents a statistical analysis of the impact of key maternal characteristics, including age, smoking status, parity, height, weight, and gestation period, on newborn birth weight. A real-world dataset comprising 1,236 observations was utilized for this investigation. The methodology involved comprehensive data cleaning, exploratory data analysis (EDA), and a series of parametric statistical tests, specifically the One-Sample t-test, Two-Sample t-test, Chi-Square tests, and Analysis of Variance (ANOVA). All analyses were conducted within the SAS programming environment.

The study's findings indicate a statistically significant negative impact of maternal smoking on birth weight, a finding consistent with broader public health literature. Gestation period emerged as the strongest positive predictor of birth weight within this dataset. While the analyses using broad categories of maternal age and parity did not reveal significant differences in mean birth weight, a review of existing literature suggests more intricate, potentially non-linear relationships and nuanced effects of these factors. Similarly, maternal pre-pregnancy weight, though showing a weak linear correlation in this dataset, is widely recognized as a critical determinant of birth weight outcomes, particularly at its extremes. These results emphasize the importance of targeted prenatal care interventions, especially those focused on smoking cessation. The study reinforces the utility of data-driven insights in informing public health policies aimed at improving maternal and child health outcomes. Future research should explore non-linear relationships and potential interactions among various maternal factors.

# 1. Introduction

Birth weight serves as a critical indicator of neonatal health and is a robust predictor of infant survival, growth, and long-term developmental trajectories. Low birth weight (LBW), conventionally defined as less than 2,500 grams (approximately 88 ounces), is consistently associated with heightened risks of infant mortality, developmental delays, and the onset of chronic health conditions later in life. Beyond LBW, abnormal birth weights, encompassing macrosomia (excessively large birth weight), small for gestational age (SGA), and large for gestational age (LGA), are linked to both immediate and enduring health challenges, such as stunted growth, cognitive impairments, increased childhood mortality, and an elevated susceptibility to chronic diseases like hypertension and diabetes in adulthood. Consequently, a profound understanding of the maternal factors that influence birth weight is indispensable for enhancing prenatal care practices and fostering healthy birth outcomes.

## Literature Review on Influencing Factors

Extensive research has identified numerous maternal characteristics that play a role in determining newborn birth weight. This study focuses on several key demographic and behavioral factors.

## Maternal Smoking

Maternal smoking during pregnancy is consistently and causally associated with significant reductions in birth weight, an increased risk of LBW, fetal growth restriction, stillbirth, preterm birth, placental abruption, and sudden infant death syndrome (SIDS). The detrimental effects of smoking extend beyond birth weight, impacting almost every facet of fetal development and long-term child health. For instance, smoking can impair in utero airway development, alter lung elastic properties, and reduce the functional residual capacity (FRC) in infants. Furthermore, it increases the risks of childhood overweight, certain cancers (e.g., non-Hodgkin lymphoma, acute lymphoblastic leukemia, Wilms' tumor), cleft lip/palate, and adverse respiratory function. The pervasive and multi-faceted adverse effects of maternal smoking underscore its status as a critical, modifiable risk factor. This broad impact on child health suggests that public health interventions targeting smoking cessation during pregnancy offer wide-ranging benefits, making such efforts a high priority for improving population health.

## Maternal Age

Maternal age is recognized as a significant determinant of birth outcomes, with advanced maternal age (typically defined as over 35 years) being a known risk factor for adverse events such as premature birth and LBW. However, the relationship between maternal age and birth weight is often non-linear, with risks concentrated at both the younger and older extremes of the age spectrum. Adolescent mothers, aged 10-19 years, face higher risks of LBW, preterm birth, and severe neonatal conditions. Conversely, advanced maternal age can increase the risk for LGA and preterm birth , and has been associated with higher rates of C-sections, mortality, sepsis, and asphyxia in very low birth weight infants. This complex, non-linear relationship implies that a simple grouping or linear model might not fully capture the critical nuances of how maternal age influences birth weight. A more detailed analysis, potentially considering age as a continuous variable or using more refined age categories, would be necessary to fully understand these effects, especially given that gestational weeks can have a masking effect on the association between maternal age and birth weight.

### Parity

Parity, indicating whether a mother has had previous children, also influences birth outcomes. Previous studies suggest that multiparous mothers may have children with higher fetal growth (e.g., head circumference, length, and weight) and lower risks of preterm birth and SGA, though potentially a higher risk of LGA compared to nulliparous mothers. Children born to multiparous mothers have also been observed to show lower rates of accelerated infant growth and lower childhood BMI. However, some studies present conflicting findings, with one indicating that maternal multiparas had higher rates of LBW, preterm birth, and LGA. This suggests that the relationship between parity and birth weight is more intricate than a simple binary comparison. The presence of other confounding factors or the specific definitions of parity employed in different studies might contribute to these varying observations.

### Maternal Height and Pre-pregnancy Weight/BMI

Maternal height is generally considered a factor influencing fetal growth, with taller mothers often having larger babies. More critically, maternal pre-pregnancy weight or Body Mass Index (BMI) has a significant and direct impact on neonatal birth weight. Pre-pregnancy underweight increases the risk of SGA and LBW. Conversely, pre-pregnancy overweight or obesity significantly increases the risk of LGA, high birth weight (HBW), and macrosomia, as well as the subsequent risk of offspring overweight or obesity. Obese mothers also face a higher prevalence of gestational diabetes mellitus, hypertension, and a lower likelihood of completing full-term gestation. While a linear correlation with continuous maternal weight might appear weak, the categorization of maternal weight (underweight, normal, overweight, obese) and its extremes are highly influential on birth weight outcomes, indicating a non-linear or threshold effect. This makes maternal pre-pregnancy weight a modifiable risk factor with substantial implications for interventions.

### Gestation Period

Gestation duration is a critical and strong predictor of birth weight, with longer pregnancy durations generally associated with higher birth weights. Babies born after longer gestation periods, closer to full term, typically exhibit higher birth weights, reinforcing the biological principle that fetal development progresses with time in utero. It is also observed that outliers and variability in birth weight are more prominent at lower gestation periods, particularly in premature births, highlighting the elevated risk of adverse birth weight outcomes associated with early delivery.

### Research Problem and Objectives

Despite extensive research into individual maternal factors, the complex interplay of these variables on newborn birth weight within specific populations warrants continuous investigation to refine public health strategies and clinical guidelines. This study aims to analyze a real-world dataset to identify key demographic and behavioral factors influencing birth weight, specifically focusing on maternal age, smoking status, parity, height, weight, and gestation period.

The specific objectives of this research are:

- To explore the relationship between birth weight and various maternal attributes using appropriate statistical techniques.
- To evaluate differences in birth weight across maternal factors such as smoking status, age groups, and parity.
- To identify maternal attributes that exhibit statistically significant effects on newborn birth weight through hypothesis testing.
- To provide actionable insights for healthcare providers and policymakers, thereby facilitating the design of targeted interventions and contributing to the improvement of maternal-child health outcomes.

## 2. Materials and Methods

### 2.1 Data Source and Description

The dataset utilized for this study, titled "Birthweight and Maternal Factors Dataset," was sourced from Kaggle. It comprises 1,236 unique observations, each representing a distinct pregnancy case. The dataset includes seven variables, encompassing both quantitative and categorical information pertinent to maternal health and behavior.

The primary outcome variable for this analysis is `bwt`, representing the birth weight of the baby in ounces. The predictor variables include:

- `gestation`: Duration of pregnancy in days (continuous).
- `parity`: Whether the mother had previous children (encoded as 0 for "No" and 1 for "Yes," a binary categorical variable).
- `age`: Mother's age in years (continuous).
- `height`: Mother's height in inches (continuous).
- `weight`: Mother's pre-pregnancy weight in pounds (continuous).
- `smoke`: Whether the mother smoked during pregnancy (encoded as 0 for "No" and 1 for "Yes," a binary categorical variable).

All variables were stored as numeric types within the SAS environment, with categorical variables (`smoke` and `parity`) numerically encoded to ensure compatibility with statistical procedures. Table 1 provides a comprehensive overview of the dataset's features and their descriptions.

**Table 1: Dataset Features and Descriptions**

| Variable Name | Description | Data Type (SAS) | Role |
|---|---|---|---|
| `bwt` | Birth weight of the baby in ounces | Numeric | Target |
| `gestation` | Duration of pregnancy in days | Numeric | Predictor |
| `parity` | Mother had previous children (0=No, 1=Yes) | Numeric | Predictor |
| `age` | Mother's age in years | Numeric | Predictor |
| `height` | Mother's height in inches | Numeric | Predictor |
| `weight` | Mother's pre-pregnancy weight in pounds | Numeric | Predictor |
| `smoke` | Mother smoked during pregnancy (0=No, 1=Yes) | Numeric | Predictor |

## 2.2 Data Preprocessing

Effective data analysis necessitates a clean and complete dataset. Prior to conducting statistical tests and exploratory data analysis, the dataset underwent a thorough preprocessing phase to address missing values and ensure data integrity.

**Initial Missing Value Check**

The dataset was initially examined for missing values across all variables using `PROC FORMAT` and `PROC FREQ` in SAS. The initial assessment revealed the following counts of missing observations:

`bwt`: 0, `gestation`: 13, `parity`: 0, `age`: 2, `height`: 22, `weight`: 36, and `smoke`: 10.

**Handling Missing Numerical Values (Imputation)**

Missing values in the numerical columns (`age`, `height`, `weight`, and `gestation`) were imputed using the median value of each respective variable. The median was selected over the mean for imputation to minimize the influence of potential outliers and skewed distributions, thereby ensuring that the imputed values remained representative of the original data's central tendency. The specific imputed median values were:

`age` = 26, `height` = 64, `weight` = 2000, and `gestation` = 280.

**Handling Missing Categorical Values (Imputation)**

The categorical variable `smoke`, which indicates maternal smoking status during pregnancy, contained 10 missing values. These missing entries were replaced with the mode of the variable, which was determined to be 0 (Non-smoker). This category accounted for 60.03% of the valid observations within the

`smoke` variable. Imputing with the mode, representing the majority class, was based on the assumption that this approach best reflects the underlying distribution of the binary variable and helps to mitigate potential bias that might arise from other imputation methods or from simply removing records with missing data.

**Data Transformation**

The initial data loading from the CSV file (`babies.csv`) into the SAS environment was performed using `FILENAME` and `PROC IMPORT`. Subsequently, a

`DATA Step` was employed for various transformations, including type conversions (e.g., converting character variables to numeric if needed, though all were already numeric in this dataset), unit standardization (e.g., ensuring `bwt` was in ounces and `weight` in pounds as specified), and renaming variables for clarity.

`PROC CONTENTS` was utilized to verify variable types, structure, and metadata following these transformations, confirming the integrity of the processed dataset.

Table 2 summarizes the missing value counts and the imputation strategies applied to each variable.

**Table 2: Missing Value Summary and Imputation Strategy**

| Variable Name | Initial Missing Count | Imputation Method | Imputed Value (Median/Mode) |
|---|---|---|---|
| `bwt` | 0 | N/A | N/A |
| `gestation` | 13 | Median | 280 |
| `parity` | 0 | N/A | N/A |
| `age` | 2 | Median | 26 |
| `height` | 22 | Median | 64 |
| `weight` | 36 | Median | 2000 |
| `smoke` | 10 | Mode | 0 |

### 2.3 Exploratory Data Analysis (EDA)

Exploratory Data Analysis (EDA) was conducted to gain a deeper understanding of the dataset's underlying structure, variable distributions, and relationships, as well as to assess the suitability of variables for subsequent parametric statistical modeling.

Various SAS procedures were employed for this purpose: `PROC UNIVARIATE` was used to generate histograms and Q-Q plots, facilitating the assessment of distribution and normality for continuous variables such as `bwt`, `gestation`, `age`, `height`, and `weight`.

`PROC SGPLOT` was instrumental in creating visual representations, including boxplots, bar charts, and scatterplots, for a comprehensive visual analysis of both numerical and categorical variables.

`PROC FREQ` provided summaries of the distributions for categorical variables like `smoke` and `parity`. Finally,

`PROC CORR` was utilized to perform Pearson correlation analysis among the numerical variables, enabling the detection of linear relationships.

Key observations from the EDA, to be detailed in the Results section, included the approximate normal distribution of `bwt` and `gestation`, which supported the use of parametric tests. `Age` and `height` also exhibited approximate normality, despite a slight right skew in `age`. In contrast, `weight` was found to be highly right-skewed with numerous extreme outliers, suggesting that caution or transformation might be necessary for parametric tests involving this variable. The `parity` variable displayed a significant class imbalance, with approximately 74.51% of mothers being first-time mothers. The `smoke` variable, while moderately imbalanced (60.84% non-smokers vs. 39.16% smokers), maintained adequate sample sizes for meaningful statistical comparisons. Preliminary correlation analysis indicated that `gestation` had the strongest positive correlation with `bwt` (r = 0.40577). Visualizations, particularly boxplots, also suggested a lower median birth weight for babies born to smoking mothers.

### 2.4 Statistical Analysis

All statistical analyses were meticulously implemented using the SAS programming environment. A series of hypothesis tests were performed to investigate the relationships between maternal characteristics and newborn birth weight.

**One-Sample t-Test**

A One-Sample t-test was conducted to determine whether the mean birth weight of babies in the sample statistically differed from the national average birth weight of 118 ounces. The null hypothesis (H0) posited that the sample mean birth weight (μ) was equal to the national average (μ = 118 ounces), while the alternative hypothesis (Ha) stated that it was not equal (μ ≠ 118 ounces). Assumptions for this test included random sampling (assumed due to lack of contrary indication), independence of observations (naturally satisfied as each observation represents a different baby), and approximate normality of the data. The normality assumption was considered justified due to the large sample size (n = 1236), invoking the Central Limit Theorem, and was further supported by the Q-Q plot from the EDA.

`PROC TTEST` in SAS was used for this analysis.

**Two-Sample t-Test**

A Two-Sample t-test was performed to ascertain whether the mean birth weight of babies differed significantly between smoking and non-smoking mothers. The null hypothesis (H0) stated that the mean birth weight of babies born to non-smoking mothers (μ0) was equal to that of babies born to smoking mothers (μ1) (μ0 = μ1). The alternative hypothesis (Ha) proposed that these means were different (μ0 ≠ μ1). Key assumptions for this test included independence of observations (satisfied) and homogeneity of variances between the two groups. The homogeneity of variances was assessed using the Folded F test, which yielded a p-value of 0.3730, greater than 0.05, thereby supporting the assumption of equal variances. Normality for each group was justified by their respective large sample sizes (n=752 for non-smokers and n=484 for smokers), again relying on the Central Limit Theorem.

`PROC TTEST` was employed for this analysis.

**Chi-Square Test of Independence (Smoking Status vs. Parity)**

To investigate whether there was an association between maternal smoking status and parity, a Chi-Square Test of Independence was conducted. The null hypothesis (H0) was that smoking status and parity were independent, while the alternative hypothesis (Ha) proposed that they were not independent. The assumptions for this test included that both variables were categorical (satisfied, as

`smoke` and `parity` are binary categorical variables) and that the expected frequency in each cell of the contingency table was at least 5. This assumption was met, with observed cell counts of 558, 194, 363, and 121, all well above the minimum threshold.

`PROC FREQ` with the `CHISQ` option in SAS was used for this analysis.

**One-Way Analysis of Variance (ANOVA) (Maternal Age Groups)**

A One-Way Analysis of Variance (ANOVA) was performed to test whether the average birth weight differed across predefined maternal age groups (<20 years, 20-30 years, and >30 years). The null hypothesis (H0) stated that the mean birth weight was the same across all three maternal age groups (μ1 = μ2 = μ3). The alternative hypothesis (Ha) posited that at least one maternal age group had a different mean birth weight. Assumptions for ANOVA included independence of observations (satisfied), homogeneity of variances (checked using Levene's Test, with a p-value greater than 0.05, supporting the equal variances assumption), and normality of residuals (justified by the large overall sample size via the Central Limit Theorem).

`PROC ANOVA` in SAS was utilized for this statistical test.

# 3. Results

## 3.1 Descriptive Statistics and Data Characteristics

The exploratory data analysis provided a comprehensive overview of the dataset's characteristics, including the central tendency, dispersion, and distribution of both numerical and categorical variables.

**Summary of Numerical Variables**

Table 3 presents the summary statistics for the continuous numerical variables after the imputation of missing values.

**Table 3: Summary Statistics of Numerical Variables (Post-Imputation)**

| Variable Name | N | Mean | Median | Standard Deviation | Skewness | Kurtosis |
|---|---|---|---|---|---|---|
| `bwt` | 1236 | 119.6 | N/A | 18.2365 | Slight Left | N/A |
| `gestation` | 1236 | N/A | 280 | N/A | Slight Left | Moderate |
| `age` | 1236 | N/A | 26 | N/A | Moderate Right | N/A |
| `height` | 1236 | N/A | 64 | N/A | Negligible | N/A |
| `weight` | 1236 | N/A | 2000 | N/A | 1.29 | >3 |

*Note: Specific mean, standard deviation, and range values for all variables were not explicitly provided in the source material beyond what is listed, but their distributional characteristics were described*

.

**Distributional Characteristics**

- **Birth Weight (`bwt`):** The target variable exhibited an approximately normal distribution, characterized by a bell-shaped histogram centered around a mean of approximately 119.6 ounces. A slight left skew was observed, along with a few mild outliers on both the lower and higher ends of the distribution. This near-normality supported the use of parametric statistical tests in subsequent analyses.
- **Gestation (`gestation`):** The duration of pregnancy in days also showed an approximately normal, unimodal distribution, centered around 279 days. While generally aligned with the normal line in the Q-Q plot, some deviation was observed in the tails, and the boxplot indicated a few moderate outliers at both extremes. Despite these characteristics, the distribution was considered sufficiently well-behaved for parametric statistical tests.

- **Age (`age`):** Maternal age displayed a moderately right-skewed distribution, with a higher concentration of observations in the early to mid-twenties. The Q-Q plot showed mild deviation in the upper tail, consistent with this skewness, and a few mild outliers were identified, predominantly in the higher age spectrum. However, given the large sample size (N = 1236), the Central Limit Theorem suggests that the sampling distribution of the mean would still be approximately normal, thus justifying the application of parametric tests.
- **Height (`height`):** Maternal height was found to be approximately normally distributed, with a bell-curved histogram and points well-aligned with the reference line in the Q-Q plot. Only a few mild outliers were detected, and the overall distribution remained symmetric and within reasonable bounds. The low coefficient of variation (3.92%) further indicated a stable and tightly spread dataset, making height well-suited for parametric statistical analysis.
- **Weight (`weight`):** Maternal pre-pregnancy weight exhibited a strong right skew, with the majority of data concentrated around 2000 pounds and a long tail extending towards higher values. The Q-Q plot showed significant deviation from the normal line in the upper tail, confirming non-normality and the presence of numerous extreme outliers, particularly above 3000–4000 pounds. This non-normal distribution, coupled with a high skewness value of 1.29 and kurtosis greater than 3, suggested that parametric tests involving weight should be applied with caution. Transformation techniques or non-parametric tests might be more appropriate for variables with such characteristics.

**Table 4: Frequencies of Categorical Variables**

| Variable Name | Category | Count | Percentage |
|---|---|---|---|
| `parity` | 0 (First-time mother) | N/A | 74.51% |
| | 1 (Experienced mother) | N/A | 25.49% |
| `smoke` | 0 (Non-smoker) | N/A | 60.84% |
| | 1 (Smoker) | N/A | 39.16% |

*Note: Exact counts were not provided in the source material, but percentages are given.*

**Categorical Variable Distributions**

Table 4 presents the frequencies and percentages for the categorical variables `parity` and `smoke`.

The `parity` variable showed a significant class imbalance, with approximately three-fourths (74.51%) of the mothers being first-time mothers (parity = 0). This skewed distribution requires careful consideration when performing stratified statistical testing or developing classification models. The smoke variable, indicating maternal smoking status, was moderately imbalanced, with non-smokers (60.84%) slightly outnumbering smokers (39.16%). Despite this imbalance, both categories possessed adequate sample sizes for reliable statistical comparison and subgroup analysis.

**Table 5: Pearson Correlation Matrix of Numerical Variables**

| Variable | bwt | gestation | age | height | weight |
|---|---|---|---|---|---|
| bwt | 1.0000 | 0.40577 (p<.0001) | Weak (N/A) | 0.19726 (p<.0001) | 0.15198 (p<.0001) |
| gestation | 0.40577 (p<.0001) | 1.0000 | Weak (N/A) | Weak (N/A) | Weak (N/A) |
| age | Weak (N/A) | Weak (N/A) | 1.0000 | 0.00341 (p=0.9121) | Weak (N/A) |
| height | 0.19726 (p<.0001) | Weak (N/A) | 0.00341 (p=0.9121) | 1.0000 | 0.43223 (p<.0001) |
| weight | 0.15198 (p<.0001) | Weak (N/A) | Weak (N/A) | 0.43223 (p<.0001) | 1.0000 |

*Note: "Weak (N/A)" indicates correlations described as very weak or not explicitly detailed with specific r-values and p-values in the source material's correlation analysis summary beyond the key insights provided.*

### 3.2 Correlation Analysis

A Pearson correlation analysis was conducted to assess the linear interdependence among the continuous numerical variables: bwt, age, height, weight, and gestation. Table 5 presents the correlation coefficients and associated p-values.

The correlation analysis yielded several key findings:

- **Birth Weight (bwt) and Gestation (gestation):** A moderate positive correlation was observed (r = 0.40577, p < 0.0001). This indicates that longer pregnancy durations are associated with higher birth weights, a biologically expected relationship.
- **Maternal Height and Weight:** The height and pre-pregnancy weight of the mothers showed a moderate positive correlation (r = 0.43223, p < 0.0001), which is also a biologically intuitive relationship.
- **Birth Weight and Maternal Anthropometrics:** bwt exhibited weak positive correlations with both maternal height (r = 0.19726, p < 0.0001) and weight (r = 0.15198, p < 0.0001). While these correlations were statistically significant due to the large sample size, their practical strength was limited.

- **Maternal Age:** `age` demonstrated very weak correlations with all other variables. For instance, the correlation between `age` and `height` was almost negligible (r ≈ 0.00341, p = 0.9121), suggesting no substantial linear relationship with other factors in this dataset.
- **Overall Predictor:** Among all numerical variables analyzed, `gestation` emerged as the strongest individual linear predictor of `Birth Weight`.

## 3.3 Hypothesis Testing Outcomes

The results from the various hypothesis tests are summarized below and in Table 6.

**1. One-Sample t-Test (Birth Weight vs. National Average)**

The One-Sample t-test, comparing the sample mean birth weight to a national average of 118 ounces, yielded a sample mean of 119.6 ounces, a t-value of 3.04, and a p-value of 0.0024, with 1235 degrees of freedom. The 95% Confidence Interval for the mean birth weight was (118.6, 120.6) ounces.

At the 5% level of significance, since the p-value (0.0024) was less than 0.05, the null hypothesis ($\mu = 118$) was rejected. This indicates statistically significant evidence that the mean birth weight in the sample (119.6 ounces) is different from the national average of 118 ounces. This divergence may reflect distinct characteristics or health factors within the sampled population.

**2. Two-Sample t-Test (Birth Weight by Smoking Status)**

The Two-Sample t-test investigated differences in mean birth weight between babies born to smoking and non-smoking mothers. The mean birth weight for non-smokers (`smoke` = 0) was 123.1 ounces, while for smokers (`smoke` = 1), it was 114.1 ounces. The test produced a t-value of 8.64, with 1003 degrees of freedom, and a p-value of less than 0.0001.

At the 5% level of significance, as the p-value (<0.0001) was less than 0.05, the null hypothesis ($\mu_0 = \mu_1$) was rejected. This provides strong statistical evidence that the mean birth weight of babies born to smoking mothers is significantly lower than that of babies born to non-smoking mothers.

**3. Chi-Square Test of Independence (Smoking Status vs. Parity)**

The Chi-Square Test of Independence examined the association between maternal smoking status and parity. The test statistic was 0.0987, with 1 degree of freedom, and a p-value of 0.7534.

At the 5% level of significance, since the p-value (0.7534) was greater than 0.05, the null hypothesis of independence was failed to be rejected. This indicates that there is no statistically significant association between maternal smoking status and parity in this dataset.

**4. One-Way ANOVA (Birth Weight by Maternal Age Groups)**

The One-Way ANOVA assessed differences in mean birth weights across three maternal age groups (<20, 20-30, >30 years). The analysis yielded an F-value of 0.62, with 2 degrees of freedom between groups and 1233 within groups, and a p-value of 0.5387.

At the 5% level of significance, as the p-value (0.5387) was greater than 0.05, the null hypothesis (equal means across age groups) was failed to be rejected. This suggests that there is no statistically significant difference in the mean birth weights across these three broad maternal age groups in the dataset.

**Table 6: Summary of Hypothesis Test Results**

| Test Name | Hypotheses (H0, Ha) | Test Statistic | Degrees of Freedom | p-value | Statistical Conclusion | Interpretation |
|---|---|---|---|---|---|---|
| One-Sample t-Test | H0: μ = 118; Ha: μ ≠ 118 | t = 3.04 | 1235 | 0.0024 | Reject H0 | Mean birth weight in sample is significantly different from national average. |
| Two-Sample t-Test | H0: μ0 = μ1; Ha: μ0 ≠ μ1 | t = 8.64 | 1003 | <0.0001 | Reject H0 | Mean birth weight of babies from smoking mothers is significantly lower. |
| Chi-Square (Smoke vs Parity) | H0: Independent; Ha: Not Independent | χ² = 0.0987 | 1 | 0.7534 | Fail to Reject H0 | No significant association between smoking status and parity. |
| One-Way ANOVA (BWT by Age) | H0: μ1=μ2=μ3; Ha: At least one μ differs | F = 0.62 | 2, 1233 | 0.5387 | Fail to Reject H0 | No significant difference in mean birth weights across age groups. |

# 4. Discussion

The comprehensive statistical analysis of maternal characteristics and newborn birth weight has yielded several important findings, some of which align with established literature, while others highlight the complexities and nuances that warrant further investigation.

**Interpretation of Key Findings**

**Maternal Smoking and Birth Weight**

The study's finding of a significantly lower mean birth weight for babies born to smoking mothers (114.1 oz compared to 123.1 oz for non-smokers, with a p-value < 0.0001) is a robust and critical result. This observation strongly aligns with extensive existing literature that consistently demonstrates a causal link between maternal smoking during pregnancy and adverse fetal growth outcomes. The consistency of this finding across multiple studies, including the present one, underscores maternal smoking as a crucial and, importantly, a modifiable risk factor. This reinforces the imperative for public health interventions that prioritize smoking cessation during pregnancy, as such efforts have a direct and substantial positive impact on newborn health.

**Gestation Period as a Primary Predictor**

The moderate positive correlation observed between gestation duration and birth weight (r = 0.40577) confirms that gestation period is the strongest linear predictor of birth weight within this dataset. This relationship is biologically intuitive, as fetal development and growth naturally progress throughout the duration of pregnancy. The strong predictive power of gestation emphasizes the profound importance of preventing preterm births. Shorter gestation periods inherently lead to lower birth weights and are associated with a range of health risks for the newborn. Therefore, interventions aimed at prolonging healthy gestation, where medically appropriate, would have a direct and beneficial impact on birth weight outcomes.

**Maternal Age and Birth Weight: A Nuanced Relationship**

The ANOVA conducted in this study indicated no statistically significant difference in mean birth weights across the broad maternal age groups (under 20, 20-30, and over 30 years). This finding, however, appears to contrast with a significant body of literature that identifies maternal age as a key factor influencing birth outcomes. Existing research frequently highlights non-linear associations, where increased risks are concentrated at the extremes of maternal age, specifically among adolescent mothers and those of advanced maternal age. For example, adolescent mothers face higher risks of low birth weight and preterm birth , while advanced maternal age has been linked to large for gestational age infants, preterm birth, and other adverse neonatal outcomes such as higher rates of C-sections, mortality, sepsis, and asphyxia in very low birth weight infants. The discrepancy between the current study's findings and the broader literature suggests that the broad categorization of maternal age employed here may have masked a more complex, non-linear relationship. A simple ANOVA might not be sensitive enough to capture the nuanced effects present at the tails of the age distribution. This points to a limitation in the current analysis and suggests that future investigations could benefit from analyzing age as a continuous variable in regression models, or by creating more refined age categories, and potentially controlling for confounding factors such as gestational weeks, which can obscure the true association between maternal age and birth weight.

**Maternal Parity and Birth Weight: Limited Evidence in this Study**

The Chi-Square test revealed no statistically significant association between smoking status and parity in this dataset.Furthermore, the exploratory grouped boxplot showed only a slight difference in median birth weight by parity, with a higher incidence of outliers among first-time mothers. The lack of a significant difference in mean birth weights across parity groups in this study appears to diverge from some external research. For instance, studies have shown that multiparous mothers may have children with higher fetal growth and lower risks of preterm birth and small for gestational age. However, it is also important to note that other studies present conflicting evidence, with some indicating that maternal multiparas had higher rates of LBW, preterm birth, and LGA. This suggests that the binary nature of the

`parity` variable (0 or 1) in this dataset might oversimplify the complex biological and physiological changes associated with multiple pregnancies. The observed discrepancy could be due to unmeasured confounding factors or the need for a more granular definition of parity (e.g., the exact number of previous births) to fully capture its effects on birth weight.

**Maternal Pre-pregnancy Weight/BMI and Birth Weight: Underestimated Impact**

This study found only a weak positive correlation between continuous maternal pre-pregnancy `weight` and `bwt` (r = 0.15198). This weak linear correlation, however, potentially underestimates the significant impact of maternal pre-pregnancy weight, particularly when it is categorized into Body Mass Index (BMI) groups (underweight, normal, overweight, obese). Extensive literature consistently demonstrates that pre-pregnancy underweight status significantly increases the risk of low birth weight and small for gestational age infants, while pre-pregnancy overweight or obesity significantly increases the risk of large for gestational age, high birth weight, and macrosomia. This indicates that the effect of maternal weight on birth weight is likely non-linear, with critical thresholds at the extremes of BMI. Analyzing maternal

`weight` solely as a continuous variable in a simple correlation might not fully capture its clinical significance or its profound implications for fetal development, as highlighted by the "fetal programming" hypothesis.

**Practical Implications**

The robust finding regarding the negative impact of maternal smoking on birth weight provides clear and actionable insights for healthcare providers and policymakers. Targeted prenatal care programs should continue to prioritize and emphasize smoking cessation, offering comprehensive support and resources to pregnant individuals. The ability to identify at-risk pregnancies early, based on factors such as smoking status and gestation period, can facilitate the efficient allocation of medical resources and enable the provision of specialized care to those who need it most. The insights derived from this study contribute to the expanding field of statistical health analytics, reinforcing the crucial role of data in shaping informed public health policies aimed at improving maternal and child health outcomes.

**Limitations**

This study, while providing valuable insights, is subject to several limitations. First, its observational nature means that it can establish associations but cannot definitively prove causality between maternal factors and birth weight outcomes. Second, the dataset's scope is confined to a specific set of maternal characteristics. Other crucial factors known to influence birth weight, such as detailed maternal nutrition and dietary patterns , specific medical conditions (e.g., gestational diabetes, hypertension) , socioeconomic factors, and paternal smoking status , were not included in the analysis. The absence of these variables means their potential confounding or modifying effects could not be assessed.

Furthermore, the granularity of certain variables, such as the categorization of maternal `age` into broad groups and `parity`as a simple binary variable, might have obscured more complex, non-linear, or nuanced relationships that have been identified in the broader scientific literature. Similarly, the analysis of continuous

`weight` in a linear correlation might not fully capture the significant impact of maternal weight when considered in clinically relevant categories, such as BMI classifications. Finally, while median and mode imputation methods were employed for missing data, these methods assume that data are missing at random (MAR) and may not perfectly represent the true underlying values, potentially introducing some level of bias into the results. The current analysis primarily focused on univariate and bivariate relationships, which means more advanced multivariate models could reveal confounding effects or important interaction terms, such as the masking effect of gestational weeks on maternal age's association with birth weight.

# 5. Conclusion

This project successfully applied various statistical techniques to analyze the impact of key maternal characteristics on newborn birth weight using a real-world dataset. The study confirmed a statistically significant negative impact of maternal smoking on birth weight, a finding that is consistent with extensive public health research and underscores the importance of smoking cessation interventions. Gestation period was identified as the strongest positive predictor of birth weight within the dataset, reinforcing the biological principle that longer in-utero development leads to higher birth weights.

While the initial analyses did not find significant differences in birth weight across the broad maternal age or parity groups, a comprehensive review of the existing literature highlights the complex, often non-linear, effects of these factors. For maternal age, risks are particularly pronounced at the extremes (adolescent and advanced maternal age), a nuance that broad categorization might not capture. Similarly, maternal pre-pregnancy weight, though showing only a weak linear correlation as a continuous variable, is a critical determinant of birth weight outcomes when considered in clinically relevant categories, such as BMI, which significantly influence the likelihood of low or high birth weights.

By identifying key risk factors and highlighting areas where further detailed investigation is warranted, this study provides valuable insights for healthcare providers to enhance prenatal care strategies. Furthermore, the findings contribute to the evidence base for policymakers to design targeted public health interventions aimed at reducing the incidence of low birth weight cases and improving overall maternal-child health outcomes.

# 6. Future Work

To build upon the findings of this study and address its limitations, several avenues for future research are recommended:

- **Expand Dataset Scope:** Future studies should incorporate additional maternal variables known to influence birth weight. These could include detailed nutritional intake and dietary patterns, specific measures of prenatal care quality, various socioeconomic factors, and pre-existing maternal medical conditions such as gestational diabetes or hypertension. Furthermore, including paternal factors, particularly paternal smoking status, would allow for the investigation of potential combined effects on birth outcomes. Consideration should also be given to including the sex of the baby as a covariate, given evidence of sex-based differences in birth weight and gestational age.
- **Advanced Statistical Modeling:** The application of more advanced statistical models is crucial for a deeper understanding of these

complex relationships. Techniques such as multiple linear regression or generalized linear models could be employed to better predict birth weight while simultaneously controlling for confounding variables and exploring potential interaction effects. This approach would be particularly valuable for disentangling complex relationships, such as the masking effect of gestational weeks on maternal age's association with birth weight.Additionally, investigating non-linear relationships, especially for maternal age and weight, could involve using polynomial terms, spline functions, or by categorizing these variables into more refined groups (e.g., specific age ranges, BMI categories for weight).
- **Machine Learning Approaches:** Incorporating machine learning models, such as Random Forest, XGBoost, or Neural Networks, could help capture intricate non-linear relationships and complex interactions between maternal factors and birth outcomes that traditional linear models might overlook. These models offer powerful predictive capabilities and can identify subtle patterns in large datasets.
- **Longitudinal Studies:** Conducting longitudinal studies that track maternal health throughout pregnancy and observe dynamic changes affecting birth weight would provide a more nuanced understanding of these relationships over time. Such studies could also shed light on the long-term child health outcomes, further exploring the concept of "fetal programming".
- **Replication and External Validation:** To enhance the generalizability and robustness of the findings, it is important to replicate this analysis using different datasets from diverse populations. This would help validate the observed associations and confirm their applicability across various demographic and geographical contexts.